%% file: Multi-layered_simulation_relations_for_linear_stochastic_systems.tex
\documentclass[10pt, twocolumn, conference]{IEEEtran}   

\usepackage[utf8]{inputenc}
\usepackage[dvipsnames]{xcolor}
\usepackage{amsfonts}
\usepackage{amsmath}
\usepackage{amsbsy}
\usepackage{amssymb}
\usepackage{mathtools}
\usepackage{subfig}
\usepackage{arydshln}
\usepackage{algorithm,algorithmicx,algpseudocode}
\usepackage[mathscr]{euscript}
\usepackage{verbatim}
\usepackage{tikz}
\usetikzlibrary{arrows,automata}

\newtheorem{definition}{Definition}
\newtheorem{theorem}{Theorem}

\newtheorem{lemma}{Lemma}

\input{Commands}

\usepackage{graphicx}   

\title{Multi-layered simulation relations for linear stochastic systems} 

\author{B.C.~van Huijgevoort and S. Haesaert%
\thanks{ B.C.~van Huijgevoort and S. Haesaert are with the Department of Electrical Engineering, TU Eindhoven, The Netherlands. Email addresses:
	{\tt\small \{b.c.v.huijgevoort, s.Haesaert\}@tue.nl}%
}%
}%

\IEEEoverridecommandlockouts

\begin{document}
\maketitle
\begin{abstract}
The design of provably correct controllers for continuous-state stochastic systems crucially depends on approximate finite-state abstractions and their accuracy quantification. For this quantification, one generally uses approximate stochastic simulation relations, whose constant precision limits the achievable guarantees on the control design. This limitation especially affects higher dimensional stochastic systems and complex formal specifications. This work allows for variable precision by defining a simulation relation that contains multiple precision layers. For bi-layered simulation relations, we develop a robust dynamic programming approach 
yielding a lower bound on the satisfaction probability of 
temporal logic specifications. We illustrate the benefit of bi-layered simulation relations for linear stochastic systems in an example.  
\end{abstract}

\section{Introduction}
Stochastic difference equations are often used to model the behavior of complex systems whose uncertainty is relevant, such as autonomous vehicles, airplanes, and drones. In this work, we are interested in automatically designing controllers for which we can give guarantees on the functionality of stochastic systems with respect to temporal logic specifications such as (sequential) reach-avoid specifications. Such automatic control synthesis is often referred to as correct-by-design control synthesis. To apply these formal synthesis methods on continuous state systems, a finite-state abstraction of the original continuous-state model is commonly used \cite{belta2017formal}. 

Abstraction-based control synthesis methods work well for most stochastic systems
\cite{julius2009approximations,lahijanian2009probabilistic,lahijanian2015formal,zamani2013symbolic,zamani2014symbolic}. However, for higher dimensional systems and more complex specifications, such as specifications with a tight labeling and a long time horizon, 
we cannot synthesize controllers that yield a high satisfaction probability.
Approaches that can handle these more complex specifications such as  \cite{cauchi2019StocHy,cauchi2019efficiency} impose restrictions on the used model classes and are subject to the curse of dimensionality. On the other hand, approaches that can handle more general model classes and allow for model order reduction to mitigate the dimensionality curse yield conservative lower bounds on the satisfaction probability for this type of complex specifications.
For more general model classes, one can use approximate simulation relations  \cite{haesaert2017verification} that quantify the abstractions via both 
 \emph{probabilistic deviations} and \emph{output precision}.  
Using these simulation relations, the abstraction accuracy can be quantified with high output precision and large probabilistic deviations for tight specifications over a short horizon and with low probabilistic deviations and low output precision for long-horizon specifications.
However, as long as these methods are considering a constant simulation relation and hence a constant abstraction accuracy, they will yield conservative results for complex specifications. Instead, in this paper, we investigate varying abstraction accuracy  by layering simulation relations for tight specifications with large time horizons.

For deterministic systems there exist methods that
construct a non-uniform abstraction grid \cite{ren2019dynamic, Tazaki2010}. More specifically, they give an approximate bisimulation relation for variable precision (or dynamic) quantization and develop a method to locally refine a coarse abstraction based on the system dynamics. Furthermore, for deterministic systems there also exist methods known as multi-layered abstraction-based control synthesis. They focus on maintaining multiple abstraction layers with different precision, where they use the coarsest abstraction when possible \cite{camara2011safety, camara2011synthesis, girard2020safety, hsu2018multi}. For stochastic models, non-uniform partitioning of the state space has been introduced for the purpose of verification \cite{Soudjani2013adaptive} 
and for verification and control synthesis in the software tools FAUST$^2$ \cite{soudjani2015fau} and StocHy \cite{cauchi2019StocHy,cauchi2019efficiency}. The latter builds on interval Markov decision processes. 

In this paper, a first step is made towards 
allowing variable precision by presenting a simulation relation that contains multiple precision layers. 
For simulation relations with two layers, we 
develop a robust dynamic programming approach such that we can compute a lower bound on the satisfaction probability of complex specifications. 

In the next section, we discuss preliminaries and formulate the problem statement for a general class of nonlinear stochastic difference equations. Section \ref{sec:SimRel}, details the current constant precision method and defines the multi-layered simulation relation for variable precision.
The following section discusses dynamic programming to compute the corresponding satisfaction probability. 
The implementation of the multi-layered method for \emph{linear} time-invariant systems and an illustrative example are given in Section \ref{sec:Impl}. 

\input{probform}

\section{Multi-layered simulation relations}\label{sec:SimRel}
\input{bisim}

\section{Multi-layered Dynamic programming}\label{sec:DP}
\input{DP} 

\section{Implementation for LTI systems}\label{sec:Impl}
\input{implementation}

\subsection{Illustrative example}\label{sec:results}
\input{CaseStudy}


\bibliographystyle{plain} 
\bibliography{Sources} 

\begin{appendices}
\section{Proof of Lemma \ref{lem:epsdelBounds} and Theorem \ref{th:theorem}} \label{app:LMIs}
For the construction of the matrix inequalities in \eqref{eq:LMI_delta}, we follow  \cite{Huijgevoort2020similarity} and model the state dynamics of the abstract model \eqref{eq:modelAbsLTI} as
$\hat{x}(t+1)\! =\! A\hat{x}(t)+B\hat{u}(t)+B_w(\hat{w}_{\gamma}(t)-\gamma(t))+\beta(t)$
with disturbance $\hat{w}_\gamma \in\mathbb{W} \subseteq \mathbb{R}^p$, shift $\gamma \in \Gamma$ and deviation $\beta \in \mathsf{B}$. The disturbance is generated by a Gaussian distribution with a shifted mean, $\hat{w}_{\gamma}\sim\mathcal{N}(\gamma,I)$. The $\beta$-term pushes the next state towards the representative point of the grid cell. Based on \cite{Huijgevoort2020similarity}, we choose stochastic kernels $\mathcal{W}_{ij}$ such that the probability of event $w-\hat{w}_\gamma =0$ is large. The error dynamics conditioned on this event equal
$x_\Delta^+  = Ax_\Delta(t)+B_w\gamma_{ij}(t)-\beta(t)$,
where state $x_\Delta$ and state update $x_\Delta^+ $ are the abbreviations of $x_\Delta(k):=x(t)-\hat{x}(t)$ and $x_\Delta(t+1)$, respectively. This can be seen as a system with state $x_\Delta$, constrained input $\gamma_{ij}$ and bounded disturbance $\beta$. 

For a given deviation $\delta_{ij}$, we compute a bound on the allowable shift as
\mbox{$\gamma_{ij} \in \Gamma_{ij} := \left\{ \gamma_{ij} \in \mathbb{R}^p \mid || \gamma_{ij} || \leq r_{ij}\right\}$} and we parameterize the shift $\gamma_{ij}=F_{ij}x_\Delta$ with the matrix $F_{ij}$. 
In the exact same fashion as the proof of Theorem 11 in \cite{Huijgevoort2020similarity}, we can show that if there exists $\lambda_{ij}$ and $F_{ij}$ such that the matrix inequalities in \eqref{eq:LMI_delta} are satisfied, then the following implications also hold
\begin{align*}
   & x_\Delta^\top D x_\Delta \leq \epsilon_i^2 \implies x_\Delta^\top F_{ij}^\top F_{ij} x_\Delta \leq r_{ij}^2 \hspace{1cm}\mbox{\itshape \small (input bound) } \\
   &  x_\Delta^\top D x_\Delta \leq \epsilon_i^2 \implies (x_\Delta^+)^\top D x_\Delta^+  \leq \alpha_{ij}^2\epsilon_i^2. \hspace{.8cm}\mbox{\itshape \small (contraction) }
\end{align*}
Therefore, we satisfy the bound $\gamma_{ij}\in\Gamma_{ij}$ and the simulation relation $\mathscr{R}_i$ describes an $\alpha_{ij}$-contractive set. Hence, using Lemma 7 in \cite{Huijgevoort2020similarity}, we can conclude that there exists a kernel $\mathcal{W}_{ij}$, such that condition 2 in Def. \ref{def:HybSimRel} is satisfied. 
Since condition 1 in Def. \ref{def:HybSimRel} was already satisfied by choosing $D$ appropriately, $\hat{M} \preceq_{\boldsymbol{\epsilon}}^{\boldsymbol{\delta}} M$ holds as long as the conditions in Theorem \ref{th:theorem} are satisfied.

Concluding, since \eqref{eq:LMI_D} holds, condition 1 in Def. \ref{def:HybSimRel} is satisfied for all $i,j$. If in addition $\lambda_{ij}$ and $F_{ij}$ satisfy \eqref{eq:LMI_delta}, then there exists a kernel $\mathcal{W}_{ij}$ such that condition 2 in Def. \ref{def:HybSimRel} holds (Lemma \ref{lem:epsdelBounds}). Once this does not only hold for a specific $i,j$, but for all $i,j \in [1, \dots, N_R]$ and there exists $i\in[1,\dots,N_R]$ with $(\hat{x}_0,x_0) \in \mathscr{R}_i$, then we have $\hat{M} \preceq_{\boldsymbol{\epsilon}}^{\boldsymbol{\delta}} M$.

\end{appendices}
\end{document}

%% file: Commands.tex
\newcommand{\until}{\mathbin{\sf U}}

\newcommand{\X}{\mathbb{X}}

\newcommand{\Y}{\mathbb{Y}}

\newcommand{\idf}{\operatorname{idf}}

\newcommand{\word}{\pmb{\pi} }

\newcommand{\FSA}{\mathcal A}

\newcommand{\Bel}{{\mathbf {T}}}

%% file: probform.tex

\section{Problem formulation}\label{sec:problem}
In this work, the Borel measurable space of a set $\mathbb{X}\subset \mathbb R^n$ is denoted by $(\mathbb{X},\mathscr{B}(\mathbb{X}))$, with $\mathscr{B}(\mathbb{X})$ the Borel sets. 
A probability measure $\mathbb{P}$ over this space has realization $x \sim \mathbb{P}$, with $x \in \mathbb{X}$. 
Furthermore, a time update of a variable $x$ is interchangeably denoted by $x(t+1), x_{t+1}$ or $x^+$.

\subsection{Preliminaries}
\textbf{Model.} In this work, we consider discrete-time   
systems 
described by a stochastic difference equation
\begin{equation}
M: \begin{cases}
x(t+1) = f(x(t),u(t),w(t)) \\
y(t) = h(x(t)), \quad \forall t \in \left\{0,1,2, \dots\right\},
\end{cases}
\label{eq:model}
\end{equation} with state $x(t)\in\mathbb{X}$, input $u(t)\in\mathbb{U}$, disturbance $w(t)\in\mathbb{W}$, output $y(t)\in\mathbb{Y}$ and with measurable functions $f: \mathbb{X} \times \mathbb{U} \times \mathbb{W} \rightarrow \mathbb{X}$ and $h: \mathbb{X} \rightarrow \mathbb{Y}$. The class of all stochastic difference equations \eqref{eq:model} with the same metric output space $(\mathbb{Y},\textbf{d}_{\mathbb{Y}})$ is denoted as $\mathcal{M}_{\mathbb{Y}}$. The system is initialized with $x(0)=x_0\in \X$ and $w(t)$ is an independently and identically distributed signal with realizations $w \sim \mathbb{P}_{w}$. 

A finite path of the model is a sequence $\omega_t:= x_0, u_0, x_1, u_1, \ldots, x_t$. An infinite path is a sequence $\omega:= x_0, u_0,\ldots $. The paths start at $x_0=x(0)$ and are build up from realizations $x_{i+1}=x(i+1)$ based on \eqref{eq:model} given a state $x(i)=x_i$, input $u(i)$ and disturbance $w(i)$ for each time step $i$.
We denote the state trajectories as $\boldsymbol{x} = x_0,x_1,\dots$, with associated suffix $\boldsymbol{x}_t = x_t,x_{t+1},\dots$. The output $y_t$  contains the variables of interest for the 
performance of the system and for each state trajectory there exists a corresponding output trajectory $\boldsymbol{y} = y_0,y_1,\dots$. 
 
A control strategy is a sequence $\boldsymbol{\mu} = (\mu_0, \mu_1, \mu_2, \dots )$ of maps $\mu_i(\omega_t)\in \mathbb U$ that assigns for each finite path $\omega_t$ an input $u_t$.  The control strategy is a Markov policy if $\mu_t$ only depends on $x_t$, and it is stationary if the policies $\mu_t$ do not depend on the time index $t$. In this work, we are interested in control strategies denoted as $C$ that can be represented with finite memory, that is, policies that are either time stationary Markov policies or have a finite internal memory.
 
\textbf{Specifications.} To express reach-avoid specifications,  
we use the syntactically co-safe linear temporal logic language (scLTL) \cite{belta2017formal,kupferman2001model}. 
This language consists of atomic propositions $p_1,p_2, \dots p_N$ that are true or false. The set of atomic propositions and the corresponding alphabet are denoted by $AP=\left\{p_1,\dots,p_N\right\}$ and $\Sigma=2^{AP}$, respectively. Each letter $\pi\in\Sigma$ contains the set of atomic propositions that are true. A (possibly infinite) string of letters forms a word $\boldsymbol{\pi} = \pi_0,\pi_1,\dots$. The output trajectory $\boldsymbol{y} = y_0,y_1,\dots$ of a system \eqref{eq:model} is translated to the word $\boldsymbol{\pi} = L(y_0),L(y_1),\dots$ using labeling function $L:\mathbb{Y}\rightarrow 2^{AP}$ that translates each output to a specific letter $\pi_t = L(y_t)$. Similarly, suffices $\boldsymbol{y}_t$ 
are translated to suffix words $\boldsymbol{\pi}_t$.
By combining atomic propositions with logical operators, the language of scLTL can be defined as follows.
\begin{definition}[scLTL syntax]
An scLTL formula $\phi$ is defined over a set of atomic propositions as
\begin{equation}\label{eq:scLTLspec}
\phi ::=  p \,|\, \lnot p \,|\, \phi_1 \wedge \phi_2 \,|\, \phi_1 \lor \phi_2 \,|\, \bigcirc \phi \,|\, \phi_1  \cup \phi_2,
\end{equation} with atomic proposition $p\in AP$. \hfill\(\Box\)
\end{definition} \noindent 
The semantics of this syntax can be given for the suffices $\boldsymbol{\pi}_t$.
An atomic proposition $\word_t \models p$ holds if $p \in \pi_t$, while a negation $\word_t \models \lnot \phi$ holds if $\word_t \not\models \phi$. Furthermore, a conjunction $\word_t \models \phi_1 \wedge \phi_2$ holds if both $\word_t \models \phi_1$ and $\word_t \models \phi_2$ are true, while a disjunction $\word_t \models \phi_1 \lor \phi_2$ holds if either $\word_t \models \phi_1$ or $\word_t \models \phi_2$ is true. Also, a next statement $\word_t \models \bigcirc \phi$ holds if $\word_{t+1} \models \phi$. Finally, an until statement $\word_t \models \phi_1 \until \phi_2$ holds if there exists an $i\in\mathbb{N}$ such that $\word_{t+i} \models \phi_2$ and for all $j \in \mathbb{N}, 0 \leq j < i$ we have $\word_{t+j} \models \phi_1$.
A system satisfies a specification if 
the generated word $\boldsymbol{\pi}_0=\boldsymbol{\pi}= L(\boldsymbol{y})$ satisfies the specification, i.e.,  
$\boldsymbol{\pi}_0 \models \phi$. 

\subsection{Problem statement}
Correct-by-design control synthesis focuses on designing controller $C$, for model $M$ and specification $\phi$, such that the controlled system $M\times C$ satisfies the specification, denoted as $M\times C \models \phi$. 
For stochastic systems, 
we are interested in the satisfaction probability, denoted as $\mathbb{P}(M \times C \models \phi)$.

\textbf{Problem.} 
Given model $M$ as in \eqref{eq:model}, an scLTL specification $\phi$ and a probability $p\in[0,1],$ find a controller $C$, such that 
\begin{equation}
\mathbb{P}(M\times C \models \phi) \geq p.
\label{eq:ContrProb}
\end{equation}

%% file: bisim.tex


Consider a continuous-state model as given in \eqref{eq:model}, approximated with the following discrete-state abstract model  
\begin{equation}
	\hat{M}: \begin{cases}
		\hat{x}(t+1) = \hat{f}(\hat{x}(t),\hat{u}(t),\hat{w}(t)) \\
		\hat{y}(t) = \hat{h}(\hat{x}(t)),
	\end{cases}
	\label{eq:modelAbs}
\end{equation}with state $\hat{x}\in\mathbb{\hat{X}}$, initialized by $\hat x(0)=\hat{x}_0$ and with input $\hat{u}\in\mathbb{\hat{U}},$ output $\hat{y}\in\mathbb{Y}$ and disturbance $\hat{w} \in\mathbb{W}$. The functions $\hat{f}:\mathbb{\hat{X}} \times \mathbb{\hat{U}} \times \mathbb{W} \rightarrow \mathbb{\hat{X}}$ and $\hat{h}: \mathbb{\hat{X}} \rightarrow \mathbb{Y}$ are assumed to be measurable. Furthermore, $\hat{w}(t)$ is an independently and identically distributed signal with realizations $\hat{w} \sim \mathbb{P}_{\hat{w}}$. 

\subsection{Stochastic simulation relations}
To give guarantees on the satisfaction probability 
we need to quantify the similarity between the two models. This quantification is performed by coupling the transitions of the models. First, the control inputs $u$ and $\hat{u}$ are coupled through an interface function denoted as
\begin{equation}
\mathcal{U}_v : \mathbb{\hat{U}}\times \mathbb{\hat{X}} \times \mathbb{X} \rightarrow \mathbb{U}.
\label{eq:interface}
\end{equation} 
Next, the probability measures $\mathbb{P}_w$ and $\mathbb{P}_{\hat{w}}$ of their disturbances $w$ and $\hat{w}$ are coupled. 
\begin{definition}[Coupling probability measures]
A coupling \cite{hollander2012probability} of  probability measures $\mathbb{P}_w$ and $\mathbb{P}_{\hat{w}}$ on the same measurable space $(\mathbb{W}, \mathscr{B}(\mathbb{W}))$ is any probability measure $\mathcal{W}$ on the product measurable space $(\mathbb{W}\times \mathbb{W}, \mathscr{B}(\mathbb{W}\times \mathbb{W}))$ whose marginals are $\mathbb{P}_w$ and $\mathbb{P}_{\hat{w}}$, that is,
\begin{align*}
    \mathcal{W}(\hat{A}\times \mathbb{W}) &= \mathbb{P}_{\hat{w}}(\hat{A}) \textrm{ for all } \hat{A}\in \mathscr{B}(\mathbb{W}) \\
    \mathcal{W}(\mathbb{W} \times A) &= \mathbb{P}_{w}(A) \textrm{ for all } A\in \mathscr{B}(\mathbb{W}). 
    \tag*{\text{\(\Box\)}}
\end{align*} 
\end{definition} \noindent
More information about this state-dependent coupling and its influence on the simulation relation can be found in \cite{haesaert2017verification,Huijgevoort2020similarity}.
Consider now the resulting coupled transitions $x(t+1)$ and $\hat{x}(t+1)$ based on respectively \eqref{eq:model} and \eqref{eq:modelAbs}, a measurable interface function $\mathcal{U}_v$ \eqref{eq:interface}, and a measurable stochastic kernel $\mathcal{W}(\cdot| \hat x, x, \hat u)$. 
The combined stochastic difference equation can then be defined as
\begin{align}\label{eq:combeqwiths}
\hspace{-.5cm} 	\hat{M}\| M:	\left\{\begin{array}{ll}\begin{pmatrix}
			\hat{x}_{t+1} \\
			x_{t+1} 
		\end{pmatrix} &= \begin{pmatrix}
			\hat{f}(\hat{x}_t,\hat{u}_t,\hat{w}_t) \\
			f(x(t), \mathcal{U}_v(\hat{u}_t,\hat{x}_t,x_t),w_t)
		\end{pmatrix}\hspace{-.5cm} \\
		y_t &= h(x_t)
	\end{array}
	\right.
\end{align}
with states $(\hat{x},x)\in \mathbb{\hat{X}}\times \mathbb{X}$, input $\hat{u}\in\mathbb{\hat{U}}$, coupled disturbance $(\hat{w},w)\sim \mathcal{W}(\cdot| \hat x, x, \hat u)$ and output $y \in \mathbb{Y}$.
Furthermore, in \cite{haesaert2018robust} it is shown that for any controller for this system there exists an equivalent controller for the original system \eqref{eq:model}. 
Given this coupled stochastic difference equation, we can analyze how close the transitions are. Suppose that you are given a simulation relation $\mathscr{R}\subset \hat{\mathbb{X}}\times \mathbb{X}$, then for all states inside this relation, $(\hat{x},x) \in \mathcal{R}$ and for all inputs $\hat{u} \in \mathbb{\hat{U}}$ we can quantify a lower-bound on the probability that the next state $(\hat{x}^+, x^+)$ is also inside this simulation relation, i.e. $(\hat{x}^+, x^+) \in \mathcal{R}$.
Hence, for all states $(\hat{x},x)\in \mathscr{R}$ we require that
\begin{equation}
    \forall \hat{u} \in \mathbb{\hat{U}}: (\hat{x}^+,x^+)\in \mathscr{R}
    \label{eq:simrelCond3}
\end{equation}
has a lower-bound on its probability denoted by $1-\delta$ given the transitions in \eqref{eq:combeqwiths}. 
To quantify the similarity between the stochastic models $M$ \eqref{eq:model} and $\hat{M}$ \eqref{eq:modelAbs}, we follow \cite{haesaert2017verification,Huijgevoort2020similarity} and consider an approximate simulation relation. 
\begin{definition}[$(\epsilon,\delta)$-stochastic simulation relation] \label{def:simRel}
Let the models $M$ and $\hat{M}$ in $\mathcal{M}_\mathbb{Y}$ with metric output space $(\Y, \textbf{d}_\mathbb{Y})$, 
and the interface function $\mathscr{U}_v$ \eqref{eq:interface} be given. Suppose that there exists a  Borel measurable stochastic kernel $\mathcal{W}$ that couples $\mathbb P_w$ and $\mathbb P_{\hat w}$ and there exists a measurable relation $\mathscr{R}\subseteq \mathbb{\hat{X}}\times \mathbb{X}$ such that $(\hat{x}_0,x_0)\in\mathscr{R}$ and such that for all $(\hat{x},x)\in \mathscr{R}$, we have that
\begin{enumerate}
    \item $\textbf{d}_{\mathbb{Y}}(\hat{y},y)\leq \epsilon$ with  $\hat y=\hat h (\hat x)$ and $y=h(x)$; and
    \item with probability at least $1-\delta$ the invariance \eqref{eq:simrelCond3} holds.
\end{enumerate}
Then $\hat{M}$ is $(\epsilon,\delta)$-stochastically simulated by $M$, and this is denoted as $\hat{M} \preceq_{\epsilon}^{\delta} M$.  \hfill\(\Box\)
\end{definition} \noindent In \cite{Huijgevoort2020similarity}, it has been shown that $\epsilon$ and $\delta$ have a trade-off. Increasing $\epsilon$ decreases the achievable $\delta$ and vice versa.

\subsection{Variable precision}
Current methods define one simulation relation for the whole state space, while we 
desire a multi-layered simulation relation $\boldsymbol{\mathscr{R}}$ that switches between multiple simulation relations to allow variable precision. 
Denote the number of simulation relations by $N_R$ and denote each simulation relation as $\mathscr{R}_i$ 
with 
precision $\epsilon_{i}$. 
A representation of such a multi-layered simulation relation with two simulation relations is given in Fig. \ref{fig:SimRelModes}. 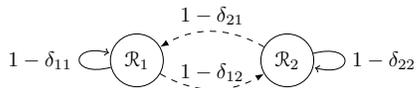
\begin{figure}[htp]
	\centering
\scalebox{0.8}{\begin{tikzpicture}[->,auto,node distance=2.5 cm,
		scale = 0.4]
		
		\node[state] (R1) {$\mathscr{R}_1$};
		\node[state] (R2) [right of = R1] {$\mathscr{R}_2$};
		
		\path[-latex] 
		(R1) edge [loop left] node [align=center]  {$1-\delta_{11}$} (R1)
		(R2) edge [loop right] node [align=center]  {$1-\delta_{22}$} (R2);
		
		\path[-latex, dashed]
		(R1) edge [bend right] node [align=center, above] {$1-\delta_{12}$}  (R2)
		(R2) edge [bend right] node [align=center, above] {$1-\delta_{21}$}  (R1);
		
	\end{tikzpicture}}
	\caption{Multi-layered simulation relation $\boldsymbol{\mathscr{R}}$ consisting of two simulation relations $\mathscr{R}_1$ and $\mathscr{R}_2$. The edges are labelled with a lower-bound on the probability that the transition occurs. }
	\label{fig:SimRelModes}
\end{figure}
Here, the self loops represent remaining in the same simulation relation, while a switch 
is indicated by the dashed arrows. Similarly to the invariance requirement in  \eqref{eq:simrelCond3}, we now associate a lower bound on the probability of each transition from $\mathscr{R}_i$ to $\mathscr{R}_j$ as $1-\delta_{ij}$.
 Furthermore, we define ${\boldsymbol{\epsilon}},{\boldsymbol {\delta}}$ with
$\boldsymbol \epsilon = \begin{matrix}(\epsilon_1, \epsilon_2, \dots, \epsilon_{N_R} \end{matrix})\mbox{ and } \boldsymbol \delta = \begin{psmallmatrix} \delta_{11}&\delta_{12}&\ldots 
	\\ \delta_{21} & \delta_{22} \\
	\vdots & & \ddots \end{psmallmatrix}.$ \smallskip

In the remainder of this paper, a switch from simulation relation $\mathscr{R}_i$ to  $\mathscr{R}_j$ is denoted by action $s_{ij}$. This assigned action determines the stochastic kernel $\mathcal{W}_{ij}$. 
Since the disturbances of the combined transitions \eqref{eq:combeqwiths} are generated from this stochastic kernel, 
\eqref{eq:combeqwiths} holds, with $(\hat w , w)\sim \mathcal W_{ij}$ if $s_t= s_{ij}$.
The input space of this combined system has been extended; that is, next to the input $u_t $ we also have a switching input $s_{t}$. Remark that for any control strategy $\boldsymbol \mu$ for $\hat{M}\| M$ there still trivially exists also a control strategy $\boldsymbol \mu'$ for $M$ that preserves the satisfaction probability. A multi-layered simulation relation is  defined as follows.
\begin{definition}[Multi-layered simulation relation] \label{def:HybSimRel}
Let the models $M$ and $\hat{M}$ in $\mathcal{M}_\mathbb{Y}$ with metric output space $(\Y, \textbf{d}_\mathbb{Y})$, and the interface function $\mathscr{U}_v$ \eqref{eq:interface} be given. If there exists measurable relations $\mathscr{R}_i\subseteq \mathbb{\hat{X}}\times \mathbb{X}$ and  Borel measurable stochastic kernels  $\mathcal{W}_{ij}$ that couple $\mathbb P_w$ and $\mathbb P_{\hat w}$ for  $i,j \in \left[1,\dots, N_R\right]^2$ 
 such that for all $i\in \left[1,\dots, N_R\right]$: 
\begin{enumerate}
\item $\forall(\hat{x},x)\in \mathscr{R}_i:\textbf{d}_{\mathbb{Y}}(\hat{y},y)\leq \epsilon_{i}$, 
\item $\forall(\hat{x},x)\in \mathscr{R}_i, \forall \hat{u} \in \mathbb{\hat{U}}: (\hat{x}^+,x^+)\in \mathscr{R}_j$ holds with probability at least $1-\delta_{ij}$ with respect to $\mathcal W_{ij}$;
\end{enumerate}
and for which there exists $i\in \left[1,\dots, N_R\right]$ with $(\hat{x}_0,x_0)\in\mathscr{R}_i$. 
Then $\hat{M}$ is stochastically simulated by $M$ in a multi-layered fashion, 
denoted as $\hat{M} \preceq_{\boldsymbol{\epsilon}}^{\boldsymbol{\delta}} M$.  \hfill\(\Box\)
\end{definition} \noindent 
This simulation relation differs from the original one in Def. \ref{def:simRel}, since it contains multiple simulation relations with different precision and therefore, allows for variable precision.

%% file: DP.tex

\subsection{scLTL satisfaction as a reachability problem}
For control synthesis purposes an scLTL specification \eqref{eq:scLTLspec} can be written as a deterministic finite-state automaton (DFA), defined by the tuple $\mathcal{A}=\left\{Q,q_0,\Sigma,\tau_\mathcal{A},Q_f\right\}$. Here, $Q$, $q_0$ and $Q_f$ denote the  set of states, initial state, and set of accepting states, respectively. Furthermore, $\Sigma=2^{AP}$ denotes the input alphabet and $\tau_\mathcal{A}:Q\times \Sigma \rightarrow Q$ is a transition function.
For any scLTL specification $\phi$ there exists a corresponding DFA $\mathcal{A}_\phi$  such that the word $\boldsymbol{\pi}$  satisfies this specification  $\boldsymbol{\pi} \models \phi$, when   $\boldsymbol{\pi}$ is accepted by $\mathcal{A}_\phi$ \cite{belta2017formal}. Here, acceptance by a DFA means that there exists a trajectory $q_0q_1q_2 \dots q_F$ with $q_F \in Q_f$ that starts with $q_0$ and evolves according to $q_{t+1} = \tau_{\mathcal{A}}(q_t,\pi_t)$.
We can therefore reason about the satisfaction of probabilistic properties over $M$ by analyzing its product composition with $\mathcal A_{\phi}$ \cite{tkachev2013quantitative} denoted as $M\otimes \mathcal A_{\phi}$. This composition yields a stochastic system with states $(x_t, q_t)\in \X\times Q$ and input $u_t$.  Given input $u_t$ the stochastic transition  from $x_t$ to $x_{t+1} $ of $M$  is represented by the transition from $(x_t,q_t)$ to $(x_{t+1},q_{t+1})$ with $q_{t+1} = \tau_{\mathcal{A}_\phi}(q_t,L( h(x_t)))$.  
Hence solving the probabilistic satisfaction  specification $\phi$ is equivalent to solving a reachability problem over  $M\otimes \mathcal A_{\phi}$ \cite{belta2017formal}. This reachability problem can be rewritten as a dynamic programming (DP) problem.

\subsection{Dynamic programming with constant precision}
Given Markov policy $\boldsymbol \mu$ for  $M\otimes \mathcal A_{\phi}$, define the time-dependent value function $V_N^{\boldsymbol \mu}$  as 
\begin{equation*}
	\textstyle   V^{\boldsymbol \mu}_N(x,q)=\mathbb E_{\boldsymbol \mu} \bigg[ \sum\limits_{ k=1}^N \mathbf{1}_{Q_f}( q_k)\! \prod \limits_{\mathclap{j = 1}}^{k-1}\mathbf{1}_{ Q \setminus Q_f}( q_j ) \bigg|( x_0,q_0)\!=\!(x,q) \bigg]
\end{equation*}
with indicator function $\textbf{1}_F(q)$ equal to $1$ if $q \in F$ and $0$ otherwise.
Since $ V^{\boldsymbol \mu}_N(x, q)$ expresses the probability that a trajectory generated by $\boldsymbol \mu$ starting from $(x,q)$ will reach the target set $Q_f$ within the time horizon $[1,\ldots,N]$, it also expresses the probability that specification $\phi$ will be satisfied in this time horizon.
Next express the associated DP operator
\begin{equation}
\Bel^u(V)({x},q) :=  \mathbb{E}_\mu \big( \max\left\{ \boldsymbol{1}_{Q_f}(q^+),V({x}^+,q^+ )\right\}\big),
\label{eq:Tmu}
\end{equation}
with $u=\mu(x,q)$ and with the implicit transitions  $q^+ =\tau_{\FSA_\phi}(q,L(h(x^+)))$. Consider a policy $\boldsymbol{\mu}_i=\cramped{(\mu_{i+1}, \ldots \mu_N)}$ with time horizon $N-i$, then it follows
that $\cramped{ V^{{\boldsymbol \mu}_{k-1}}_{{N-k}+1}} = \cramped{\Bel^{{ \mu}_k}}\cramped{( V^{{\boldsymbol \mu}_k}_{N-k})}$. Thus if $\cramped{ V^{{\boldsymbol \mu}_k}_{N-k}}$ expresses the probability of reaching $Q_f$ within $N-k$ steps, then $ \Bel^{\mu}_k \cramped{ (V^{{\boldsymbol \mu}_k}_{N-k}) }$ expresses the probability of reaching $Q_f$ within $N-k+1$ steps with policy ${\boldsymbol \mu}_{k-1}$. It follows that for a stationary policy ${\boldsymbol \mu} $, the infinite-horizon value function can be computed as $ V_\infty^{\boldsymbol \mu} = \lim_{N\rightarrow \infty} (\Bel^{\boldsymbol{\mu}})^N  V_0$ with ${V}_0 \equiv 0$. Furthermore, the optimal DP operator $\Bel^\ast(\cdot) := \sup_{\mu} \Bel^\mu(\cdot)$ can be used to compute the optimal converged value function $V^\ast_\infty$. The corresponding satisfaction probability can now be computed as $\mathbb{P}^{\boldsymbol{\mu}} := \max(\boldsymbol{1}_{Q_f}(\bar{q}_0),V^{\ast}_{\infty}(x_0,\bar{q}_0))$, with $\bar{q}_0=\tau(q_0,L(h(x_0))$. When the policy $\boldsymbol{\mu}$, or equivalently the controller $C$,  yields a satisfaction probability higher than $p$, then \eqref{eq:ContrProb} is satisfied and the synthesis problem is solved.

Due to its continuous states the DP formulation above cannot be computed for the original model $M$, so we use abstract model $\hat{M}$. Next, we
adjust the method in \cite{haesaert2018robust} to take into account the output- and probability deviations.

\subsection{Bi-layered dynamic programming approach}
\begin{figure}[htp]
	\centering
\subfloat[\mbox{ }] {\includegraphics[width=0.28\columnwidth]{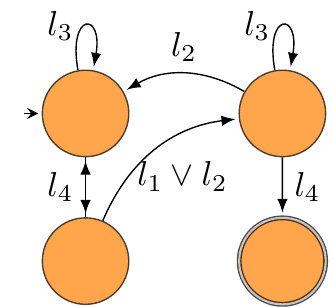}\label{fig:layereda}}\quad
\subfloat[\mbox{ }]  {\includegraphics[width=0.28\columnwidth]{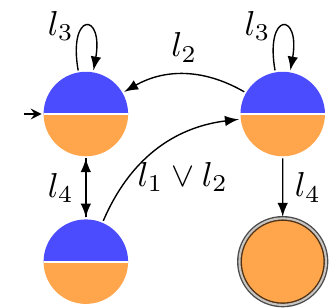}\label{fig:layeredb}}\quad 
\subfloat[\mbox{ }] {\includegraphics[width=0.28\columnwidth]{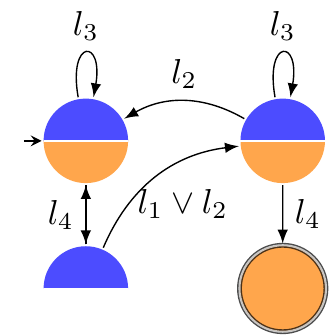}\label{fig:layeredc}}
\caption{DFA $\mathcal A_{\phi}$ with labels $l_i\in\Sigma$, (a) with discrete modes with one layer, (b) with two layers and (c) with the 
	high-precision layer only in modes with self loops.} 
\end{figure}

To implement a layered DP approach, each simulation relation $\mathscr{R}_i$ gets its own layer $i$ to which we assign 
a value function $V(x,q,i)$. In the remainder of this section, we present a bi-layered approach, 
where two simulation relations $\mathscr{R}_1$ and $\mathscr{R}_2$ with $\mathscr{R}_1\supseteq
\mathscr{R}_2$,  $\epsilon_1\geq \epsilon_2$, and $\delta_1\leq \delta_2$ are given. 
We further assume that the layers and corresponding switching strategies are given. A switching strategy consists of switching actions defined for all abstract states $\hat{x} \in \mathbb{\hat{X}}$ in each layer. 
In Fig. \ref{fig:layeredb}, we see a DFA $\mathcal{A}_\phi$ that is constructed for a bi-layered approach and with edges labeled by $l_i \in \Sigma$. Here, the fully orange modes consist of only layer $2$, while 
in the other modes both layers are created.

The value function defines a lower bound on the probability that specification $\phi$ will be satisfied in the time horizon $\left[1, \dots, N \right]$. We can now define a robust operator $\Bel^{\hat{u}}_{s_{ij}}$
as
\begin{align}
    & \Bel^{\hat{u}}_{s_{ij}}(V)(\hat{x},q,i) = \notag\\
    &\textbf{L}\big( \mathbb{E}_{\hat{u}} \big( \min_{q^+ \in Q_{\epsilon_{j}}^+} \max\left\{ \boldsymbol{1}_{Q_f}(q^+),V(\hat{x}^+,q^+,j)\right\}\big) -\delta_{ij}  \big),
    \label{eq:Tmu-epsdel}
\end{align}  with $\textbf{L}:\mathbb{R}\rightarrow [0,1]$ a truncation function defined as $\textbf{L}(\cdot):=\min(1,\max(0,\cdot))$ and with 
\begin{align}
Q_{\epsilon_{j}}^+(q,\hat{y}^+) := \left\{\tau_{\mathcal{A}} (q,  L(y^+)) \mid 
 ||y^+-\hat{y}^+|| \leq \epsilon_{j} \right\}.
\label{eq:Qplus}
\end{align} 
For a given switching policy $\mu^s:\mathbb{\hat X} \times Q \times [1,2] \rightarrow \mathbb{\hat U}\times [s_{i1}, s_{i2}]$,  we define
$\Bel^{\mu^s} (V)(\hat{x},q,i)= \Bel^{\hat{u}}_{s_{ij}}(V)(\hat{x},q,i) \mbox{ with } (\hat{u},s_{ij})=\mu^s(\hat x, q, i).$
Consider a policy $\boldsymbol{\mu}_k^s=(\mu_{k+1}^s, \dots, \mu_N^s)$,  
then for all $(\hat{x},q, i)$
we have that $V_{N-k+1}^{\boldsymbol{\mu}^s_{k-1}}= \Bel^{\mu_k^s}(V_{N-k}^{\boldsymbol{\mu}^s_k})$, initialized with ${V}_0 \equiv 0$. As before, for a stationary policy $\boldsymbol{\mu}^s $, the infinite-horizon value function for both layers 
can be computed as $ V_\infty^{\boldsymbol{\mu}^s} = \lim_{N\rightarrow \infty} (\Bel^{\boldsymbol{\mu^s}})^N  V_0$ with ${V}_0 \equiv 0$. Furthermore, the optimal robust operator $\Bel^\ast(\cdot) := \sup_{\mu^s} \Bel^{\mu^s}(\cdot)$ can be used to compute the optimal converged value function $V^\ast_\infty$.
 \\ \smallskip
Consider a control strategy $\boldsymbol \mu^s$ for $\hat M$. This strategy can also be implemented on the combined model $\hat{M}\| M$ and we can denote the value function of the combined model as $V_c(\hat{x},x,q)$. As mentioned before, the control strategy for the combined model can be refined to a control strategy of the original model $M$ \eqref{eq:model}. Although $V_c(\hat{x},x,q)$  expresses the probability of satisfaction, it cannot be computed directly, instead we can compute $V(\hat{x},q,i)$ over the abstract model $\hat M$ using \eqref{eq:Tmu-epsdel}.
\begin{lemma}\label{th:ValueFunc}
Suppose $\hat{M} \preceq_{\boldsymbol{\epsilon}}^{\boldsymbol{\delta}} M$ with a multi-layered simulation relation $\boldsymbol{\mathscr{R}}$
is given. Let $V(\hat{x},q,j) \leq V_c(\hat{x},x,q)$ for all $(\hat{x},x)\in \mathscr{R}_j$, then
\begin{equation}
    \Bel^{\hat{u}}_{s_{ij}}(V)(\hat{x},q,i) \leq \Bel^{\hat{u}}(V_{c})(\hat{x},x,q) \quad \forall (\hat{x},x) \in \mathscr{R}_i,
\end{equation} where $\Bel^{\hat{u}}_{s_{ij}}(V)(\hat{x},q,i)$ is the $(\boldsymbol{\epsilon},\boldsymbol{\delta})$-robust operator \eqref{eq:Tmu-epsdel} with respect to stochastic transitions of $\hat{M}$ and $\Bel^{\hat{u}}(V_{c})(\hat{x},x,q)$ is the exact recursion \eqref{eq:Tmu} with respect to the combined stochastic transitions \eqref{eq:combeqwiths}.   \hfill\(\Box\)
\end{lemma}
\begin{proof}The proof of Lemma \ref{th:ValueFunc} follows along the same lines of the proof of Lemma 3 
in \cite{haesaert2018robust}. 
\end{proof}

The value function gives the probability of satisfying the specification after 1 time step, by including the first time instance based on $x_0$, we can compute the robust satisfaction probability, that is
\begin{equation}
\mathbb{R}^{\boldsymbol{\mu}^s} := \max(\boldsymbol{1}_{Q_f}(\bar{q}_0),V^{\boldsymbol{\mu}^s}_{\infty}(x_0,\bar{q}_0)),
\end{equation} with $\bar{q}_0 = \tau_{\mathcal{A}_{\phi}}(q_0,L(h(x_0)))$.
The robust satisfaction probability gives a lower-bound on the actual satisfaction probability $\mathbb{P}^{\boldsymbol{\mu}^s}$. When the policy $\boldsymbol{\mu}^s$ defined by controller $C$ yields a robust satisfaction probability higher than $p$, 
then \eqref{eq:ContrProb} is satisfied and the control synthesis problem is solved.

\subsection{Bi-layered dynamic programming with partial covers}
To decrease the computation time, consider 
layer 2 to be only present in modes with a self-loop. Such a pruned bi-layered DFA is illustrated in Fig. \ref{fig:layeredc}. To decrease the computation time even further, we disregard action $s_{21}$. Such a switching strategy
is shown in Fig. \ref{fig:CaseStudyTotal}.
For layer 1 (blue), 
action $s_{11}$ and $s_{12}$ hold respectively for all states inside the blue and hatched orange region. The action for layer 2 (orange) equals $s_{22}$ until a new DFA state is reached.

To mitigate the effect of partial covers, we modify the DP iterations initialized with value functions $V_0(\hat{x},q,i) \equiv 0$ with  $i \in \left[1,2\right]$.
First, for all states $\hat{x}_k \in \mathbb{\hat{X}}$ that are not inside layer $j$, we set the value function $V_l(\hat{x}_k,q,j) = 0$ for all iterations $l$.
Since for all $(\hat x, x)\in \mathscr{R}_2$, we also have that $(\hat x, x)\in \mathscr{R}_1$ this implies that switching to layer 1 when layer 2 is missing comes for free. 
Therefore, with some abuse of notation, we define a piecewise maximum value function as $V(\hat{x}^+,q^+, \leq 2) = \max_{j\in[1,2]}  V(\hat{x}^+,q^+,j) .$
Now, the adjusted robust operator is defined as 
\begin{align}
    & \Bel^{\hat{u}}_{s_{i2}}(V)(\hat{x},q,i) = \notag\\
    &\textbf{L}\big( \mathbb{E}_{\hat{u}} \big( \min_{q^+ \in Q_{\epsilon_{2}}^+} \left\{ \boldsymbol{1}_{Q_f}(q^+),V(\hat{x}^+,q^+,\leq 2))\right\}\big) -\delta_{i2}  \big),
\end{align} with $Q_{\epsilon_{2}}$ as in \eqref{eq:Qplus}.
This adjusted operator is valid as it preserves the lower-bound defined in \eqref{eq:Tmu-epsdel} and can hence be used interchangeably.

%% file: implementation.tex

Let the models $M$ \eqref{eq:model} and $\hat{M}$ \eqref{eq:modelAbs} be linear time-invariant (LTI) systems whose behavior is described by the following stochastic difference equations 
\begin{equation}
M: \begin{cases}
x(t+1) = Ax(t)+Bu(t)+B_ww(t) \\
y(t) = Cx(t), \textrm{ and }
\end{cases}
\label{eq:modelLTI}
\end{equation}
\begin{equation}
    \hat{M}: \begin{cases}
    \hat{x}(t+1) = \Pi\left( A\hat{x}(t)+B\hat{u}(t)+B_w\hat{w}(t) \right) \\
\hat{y}(t) = C\hat{x}(t),
    \end{cases}
    \label{eq:modelAbsLTI}
\end{equation}
with matrices $A, B, B_w$ and $C$ of corresponding sizes and with the disturbances $w(t),\hat w(t)$  generated by the standard Gaussian distribution, i.e., $w(t)\sim\mathcal{N}(0,I)=\mathbb P_{w}$ and $\hat w(t)\sim\mathcal{N}(0,I)=\mathbb P_{\hat w}$. 
The abstract model is constructed by partitioning the state space $\mathbb{X}$ in a finite number of regions $\mathbb{A}_i \subset \mathbb{X}$ and operator $\Pi(\cdot): \mathbb{X}\rightarrow \mathbb{\hat{X}}$ maps states $x\in \mathbb{A}_i$ to their representative points $\hat{x}_i\in\mathbb{\hat{X}}$. We assume that the  
regions $\mathbb{A}_i$ are designed in such a way that the set $\mathsf{B}:=\left\{\Pi(x)-x \mid x\in \mathbb{X} \right\}$ is a bounded polytope and has vertices $\textrm{vert}(\mathsf{B})$. Details on constructing such an abstract LTI system can be found in \cite{haesaert2018robust}.

\subsection{Computing the multi-layered simulation relations}
To compute the multi-layered simulation relations in 
Def. \ref{def:HybSimRel},
we choose the interface function $
u(t) = \mathcal{U}_v(\hat{u}_t,\hat{x}_t,x_t) $ as 
$u(t)=\hat{u}(t)$
and consider simulation relations $\mathcal{R}_i$
\begin{equation}
	\mathcal{R}_i:=\left\{(\hat{x} , x)\in  \mathbb{\hat{X}}\times  \mathbb{X} \mid ||x-\hat{x}||_D \leq \epsilon_i \right\},
	\label{eq:simrel}
\end{equation}
where $||x||_D$ denotes the weighted two-norm, that is, $||x||_D = \sqrt{x^TDx}$  with $D$ a symmetric positive definite matrix  $D=D^T \succ 0$. We use the same weighting matrix $D$ for all simulation relations $\mathscr{R}_i$, with $i\in \left[1,2, \dots N_R \right]$. 

For these  relations, condition 1 in Def. \ref{def:HybSimRel} is satisfied by choosing weighting matrix $D \succ 0$, such that 
\begin{equation}
C^TC\preceq D.
\label{eq:LMI_D}
\end{equation}

We can now construct kernels $\mathcal W_{ij}$ in a similar way as in \cite{Huijgevoort2020similarity}. By doing so, condition 2 of Def. \ref{def:HybSimRel} can be quantified via contractive sets for the error dynamics $x_{t+1}-\hat{x}_{t+1}$ based on the combined transitions \eqref{eq:combeqwiths}.
We assume that there exists factors $\alpha_{ij}$ with $\epsilon_j = \alpha_{ij}\epsilon_i$ that represent the set contraction between the different simulation relations.
Now, we can describe the satisfaction of condition 2 as a function of $\delta_{ij}, \alpha_{ij}$ and $\epsilon_{i}$.
\begin{lemma}\label{lem:epsdelBounds}
Consider models $M$ \eqref{eq:modelLTI} and $\hat{M}$ \eqref{eq:modelAbsLTI} for which simulation relations $\mathscr{R}_i$ and $\mathscr{R}_j$ as in \eqref{eq:simrel}  are given with weighting matrix $D$ satisfying \eqref{eq:LMI_D}. For given $\delta_{ij}, \alpha_{ij},$ and $ \epsilon_i$, consider matrix inequalities \begin{subequations}\label{eq:LMI_delta}\begin{align}
& \hspace{-.5cm}\begin{bsmallmatrix}
\frac{1}{\epsilon_i^2}D & F_{ij}^T\\
F_{ij} & r_{ij}^2I
\end{bsmallmatrix}\succeq 0,\hspace{2.5cm}\mbox{\itshape \small (input bound) } \\
&  \hspace{-.5cm}\begin{bsmallmatrix}
\lambda_{ij} D & \ast & \ast   \\
0 & (\alpha_{ij}^2-\lambda_{ij})\epsilon_i^2 & \ast \\ 
D(A+B_wF_{ij}) & D\beta_l & D
\end{bsmallmatrix} \succeq 0  \mbox{  \itshape \small  (contraction) }
\end{align}\end{subequations}
parameterized with $\lambda_{ij}>0$ and with the  matrix 
$F_{ij}$ for $r_{ij} = |2\idf \left( \frac{1-\delta_{ij}}{2} \right) |$ and for all $\beta_l \in \textrm{vert}(\mathsf{B}) $. 
Here, $\idf$ denotes the inverse distribution function of the Gaussian distribution.
If there exists $\lambda_{ij}$ and $F_{ij}$ such that  the matrix inequalities in \eqref{eq:LMI_delta} are satisfied, then there exists a $\mathcal W_{ij}$ such that condition 2 in Def. \ref{def:HybSimRel} is satisfied. \hfill\(\Box\)
\end{lemma}

\begin{theorem}\label{th:theorem}
Consider models $M$ \eqref{eq:modelLTI} and $\hat{M}$ \eqref{eq:modelAbsLTI} for which simulation relations $\mathscr{R}_i$ and $\mathscr{R}_j$ as in \eqref{eq:simrel} are given with weighting matrix $D$ satisfying \eqref{eq:LMI_D}.
If the inequalities \eqref{eq:LMI_delta} hold for all $i,j \in [1,\dots, N_R]^2$ and  
there exists $i\in \left[1, \dots, N_R\right]$ with $(\hat{x}_0,x_0)\in\mathscr{R}_i$
then $\hat{M}$ is stochastically simulated by $M$ in a multi-layered fashion as in Def. \ref{def:HybSimRel}, denoted as $\hat{M}\preceq_{\boldsymbol{\epsilon}}^{\boldsymbol{\delta}} M$. \hfill\(\Box\)
\end{theorem}

\begin{proof}
The proof of both Lemma \ref{lem:epsdelBounds} and Theorem \ref{th:theorem} can be found in the appendix. 
It builds on top of the proofs of Theorem 10 and Theorem 11 in \cite{Huijgevoort2020similarity} for invariant sets. Instead of invariant sets, the proof uses contractive sets to deal with the multi-layered simulation relation. 
\end{proof}

%% file: CaseStudy.tex

\begin{figure}
\centering
\includegraphics[width=\columnwidth]{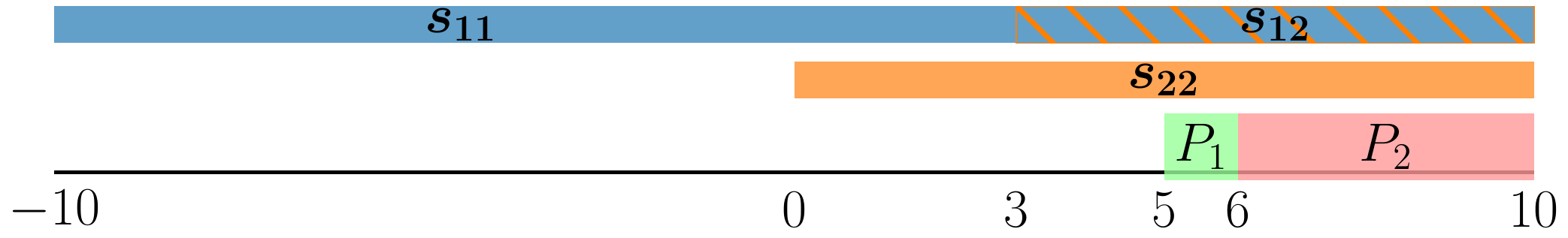}
\caption{Parking areas $P_1$, $P_2$ and switching strategy 
with blue layer $\mathscr{R}_1$ and orange layer $\mathscr{R}_2$.}
\label{fig:CaseStudyTotal}
\end{figure}

\begin{figure}
\centering
\includegraphics[width=0.6\columnwidth]{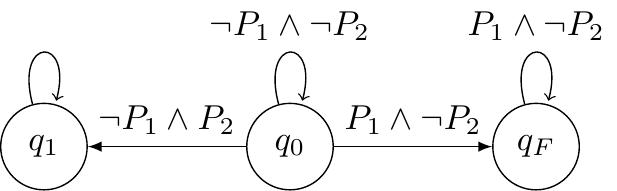}
\caption{DFA associated with specification $\phi_{park}= \lnot P_2 \until P_1$.}
\label{fig:DFA}
\end{figure}

As an illustrative example, we consider parking a car in a one-dimensional space. 
The goal of the controller is to guarantee that the car parks in the green area $P_1$, without going through the red area $P_2$, as illustrated in Fig. \ref{fig:CaseStudyTotal}. This specification can be written as $\phi_{park}= \lnot P_2 \until P_1$ and can be represented by the DFA given in Fig. \ref{fig:DFA}.
The dynamics of the car are modeled using an LTI stochastic difference equation as in \eqref{eq:modelLTI} with $A=0.9, B=0.5$ and $B_w=C=1$. We used states $x\in \mathbb{X} = [-10,10]$, inputs $u \in \mathbb{U}=[-1,1]$, outputs $y\in \mathbb{Y}=\mathbb{X}$ and Gaussian disturbance $w \sim \mathcal{N}(0,0.5)$. 
We considered the regions $P_1 = [5,6\rangle, P_2 = [6,10]$ and used the following labeling function
    \begin{equation}
        L(y) = \begin{cases}
        \lnot P_1 \land \lnot P_2 & \textrm{if } y<5 \\
        P_1 \land \lnot P_2 & \textrm{if } 5\leq y <6 \\
        \lnot P_1 \land P_2 & \textrm{if } 6\leq y \leq 10. \\
        \end{cases}
    \end{equation}

\begin{figure}
\centering
\includegraphics[width=\columnwidth]{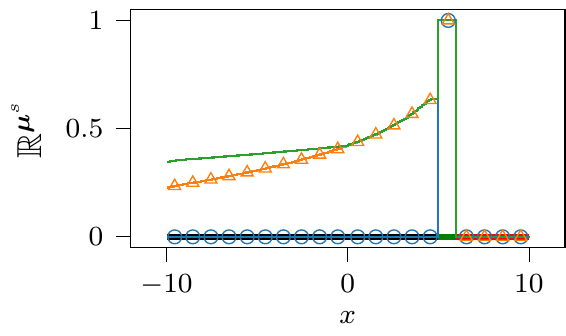}
\caption{Satisfaction probability of the example, where the blue circles and orange triangles are obtained with respectively only using 
$\mathscr{R}_1$, \mbox{$(\epsilon, \delta) = (0.5,0)$} and 
$\mathscr{R}_2$, \mbox{$(\epsilon, \delta) \approx (0.2,0.01)$}. Switching between these two simulation relations 
with switching strategy \eqref{eq:SwitchStrat} yields the green line.}
\label{fig:results}
\end{figure}

We obtained abstract model $\hat{M}$ in the form of \eqref{eq:modelAbsLTI} by partitioning 
with regions of size $0.1$
with $\mathsf{B}=[-0.05,0.05]$ and $\hat{u}\in\mathbb{\hat{U}} = \left[-1, -\frac{2}{3}, -\frac{1}{3}, \dots , 1\right]$. 
We quantified the accuracy of $\hat M$ with a bi-layered simulation relation.
The first layer with $\mathscr{R}_1$, and 
 $(\epsilon_{1}, \delta_{11}) = (0.5,0)$ covers the complete state space. 
The second layer with $\mathscr{R}_2$ has deviation $\delta_{22} =0.012$ and 
only covers $0 \leq \hat{x} \leq 10$.
We chose  $\delta_{12}=0.12$ and output precision $\epsilon_2= 0.1984$
that 
satisfy Lemma \ref{lem:epsdelBounds}.
As illustrated in Fig. \ref{fig:CaseStudyTotal}, we  chose the switching strategy:
\begin{equation} 
\mu^s = \begin{cases}
        s_{11} & \textrm{ if } -10 \leq \hat{x} \leq 3 \textrm{ and } i=1 \\
        s_{12} & \textrm{ if } 3 < \hat{x} \leq 10\textrm{ and } i=1 \\
        s_{22} & \textrm{ if } 0 < \hat{x} \leq 10\textrm{ and } i=2.
    \end{cases}
    \label{eq:SwitchStrat}
\end{equation} 
Together, this led to the satisfaction probability 
 in Fig. \ref{fig:results}. 

A constant precision with either simulation relation $\mathscr{R}_1$ or $\mathscr{R}_2$ yields the conservative satisfaction probability indicated by the respective blue circles and orange triangles in Fig \ref{fig:results}. The bi-layered method (green line) takes advantage of both simulation relations. Close to the parking areas 
simulation relation $\mathscr{R}_2$ is generally active, which compared to simulation relation $\mathscr{R}_1$ gives us a non-zero satisfaction probability.
Switching to layer 1 limits the rapid decrease of the satisfaction probability further from the parking areas, which is normally caused by the relatively high value of $\delta_{22}$. 

Concluding, the multi-layered method allows switching between multiple simulation relations and makes it possible to use the advantages of each individual simulation relation. 
Therefore, the satisfaction probability increases and is more accurate than when using constant precision.